\documentclass[prl,aps,twocolumn,amsmath,amssymb]{revtex4}
\usepackage{graphicx}
\usepackage{dcolumn}
\usepackage{bm}

\begin{document}

\preprint{APS/123-QED}
\title{Singlet-Triplet Transition Tuned by Asymmetric Gate Voltages in a Quantum Ring}

\author{A. Fuhrer,$^1$ T. Ihn,$^1$ K. Ensslin,$^1$ W. Wegscheider,$^2$ M. Bichler$^3$}

\affiliation{
$^1$Solid State Physics Laboratory, ETH Z\"urich, 8093 Z\"urich, Switzerland\\
$^2$Institut f\"ur experimentelle und angewandte Physik, Universit\"at Regensburg, Germany\\
$^3$Walter Schottky Institut, Technische Universit\"at M\"unchen, Germany}

\date{\today}

\begin{abstract}
Wavefunction and interaction effects in the addition spectrum of a Coulomb blockaded many electron quantum ring are investigated as a function of asymmetrically applied gate voltages and magnetic field. Hartree and exchange contributions to the interaction are quantitatively evaluated at a crossing between states extended around the ring and states which are more localized in one arm of the ring. A gate tunable singlet-triplet transition of the two uppermost levels of this many electron ring is identified at zero magnetic field.
\end{abstract}

\pacs{Valid PACS appear here}

\maketitle

The energy spectra of few-electron quantum dots, so called artificial atoms, are well understood~\cite{96tarucha}. Interaction effects and shell filling according to Hund's rules have been measured in good agreement with theoretical calculations.
It was shown~\cite{00tarucha} that level-crossings induced by a magnetic field may lead to singlet-triplet transitions. Comparison with a model allowed to extract exchange and direct Coulomb interaction contributions. A tunable transition from a singlet to a triplet ground state, i.e. the control of the exchange splitting relative to the single-particle splitting, is of interest  in view of possible quantum computing applications~\cite{98loss,03saraga}.
For a laterally defined few-electron dot~\cite{02kyriakidis} it was shown that such a transition may also be induced by shape deformations of the potential when tuning the gate voltage. Recently this idea was taken up in an experiment~\cite{03kogan} in the Kondo regime where indications of a singlet-triplet transition at zero magnetic field were found.

In contrast to few electron dots, energy spectra of many electron dots, as extracted from addition spectroscopy measurements in the Coulomb blockade regime, are more difficult to interpret and only a statistical analysis is possible~\cite{01ullmo,00baranger,03jiang,01luscher}. Recently, however, we have shown that in the case of a quantum ring with weak interactions, it is possible to understand in great detail the energy spectrum measured as a function of a magnetic field applied normal to the plane of the ring~\cite{03ihn,01Bfuhrer}. We used the constant interaction model to extract an experimental single-particle spectrum. 

In the present study we investigate interaction effects beyond the constant interaction in the ring spectrum. Tuning the symmetry between the arms of the ring by applying gate voltages allows to induce crossings between rather localized and more extended ring states. We can determine the magnitude of different interaction contributions in the quantum ring. This allows us to identify gate voltage regimes where the two uppermost electrons occupy either a singlet or a triplet state. At the same time we are able to characterize qualitatively the orbital wave functions associated with energy levels.

\begin{figure}[b]
\centering
\includegraphics[width=3in]{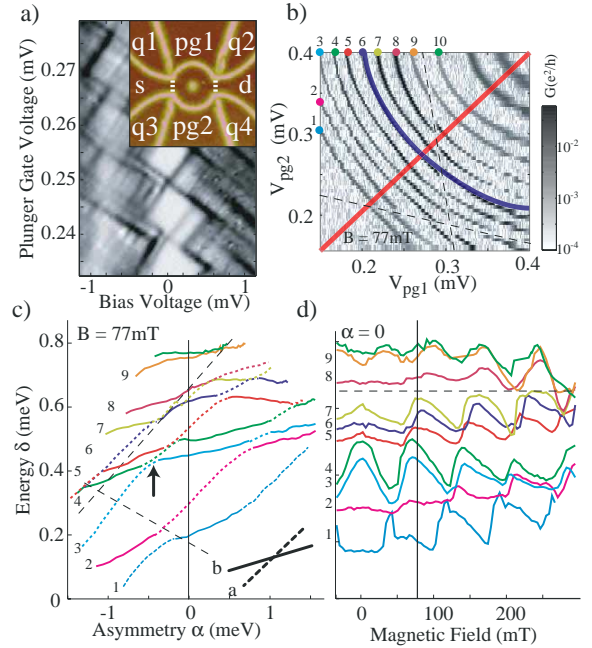}
  \caption{\small (a) inset: SFM image of the quantum ring.
  Main figure: Coulomb blockade diamonds as a function of $V_\mathrm{pg1}=V_\mathrm{pg2}$. (b) Grey scale plot of the conductance as a function of $V_\mathrm{pg1}$ and $V_\mathrm{pg2}$.  (c) Addition spectrum with a constant $E_{\mathrm C}=310~\mu$eV subtracted from peak separations. (d) Spectrum as a function of magnetic field. Dashed lines in (c) and (d) indicate parametric charge rearrangements. Lines with the same color in (c) and (d) are corresponding conductance peaks.}\label{fig1}
\end{figure}
The sample is an [Al]GaAs heterostructure containing a two-dimensional electron gas (2DEG) 34 nm below the surface. The electron sheet density is $5\times10^{15}~\mathrm{m}^{-2}$ and a mobility of $90~\mathrm{m}^2/\mathrm{Vs}$ at 4.2~K is found. The quantum ring is fabricated by local oxidation of the GaAs surface with a scanning force microscope (SFM) \cite{02fuhrer}. An SFM-image of the structure is shown in the inset of Fig.~\ref{fig1}(a). The ring has a radius of $r_0=132~\mathrm{nm}$ and is connected to source (s) and drain (d) through two tunnel barriers (dashed lines). These barriers are tuned by four in-plane gates (q1-q4) and an additional metallic top gate covering the whole structure. By applying gate voltages $V_\mathrm{pg1}=V_\mathrm{pg2}$ to the two in-plane plunger gates pg1 and pg2 the number of electrons on the ring is tuned one by one along the red line in Fig.~\ref{fig1}(b). We estimate the ring to contain more than 60 and less than 200 electrons~\cite{03ihn}. For asymmetrically applied gate voltages along the curved blue line in Fig.~\ref{fig1}(b) the electron number is constant but the symmetry of the ring is modified. 
The curvature of the Coulomb peak positions reflects the voltage dependence of the in-plane gate lever arms $\beta_{\mathrm{pg}i}\left(V_{\mathrm{pg}i}\right)$. We determine the average lever arm for each value of in-plane gate voltage from Coulomb blockade diamonds [Fig.~\ref{fig1}(a)] as a function of top and in-plane gate voltages~\cite{03ihn}.
The two plunger gates tune the energy $\delta$ of ring states according to
\[\delta\left(V_{\mathrm{pg1}},V_{\mathrm{pg2}}\right)=-e\left[\beta_{\mathrm{pg1}}\left(V_{\mathrm{pg1}}\right) V_{\mathrm{pg1}}+\beta_{\mathrm{pg2}}\left(V_{\mathrm{pg2}}\right) V_{\mathrm{pg2}}\right].\]
We define the asymmetry parameter $\alpha$ as
\[\alpha\left(V_{\mathrm{pg1}},V_{\mathrm{pg2}}\right)=-e\left[\beta_{\mathrm{pg1}}\left(V_{\mathrm{pg1}}\right) V_{\mathrm{pg1}}-\beta_{\mathrm{pg2}}\left(V_{\mathrm{pg2}}\right) V_{\mathrm{pg2}}\right].\]
This corresponds to a coordinate transformation where the blue (red) line in Fig.~\ref{fig1}(b) becomes the horizontal (vertical) axis in Figs.~\ref{fig1}(c) and (d).  

The addition spectrum in Fig.~\ref{fig1}(c) is obtained by fitting each Coulomb peak and plotting its position as a function of $\alpha$ and $\delta$. A constant $E_{\mathrm{\small C}}=310~\mu$eV, i.e. the minimum peak spacing in the parameter range investigated, was subtracted from all peak spacings~\cite{error}. The Coulomb peaks are numbered with increasing energy $\delta$ and presented in the same color throughout the paper.
The spectrum as a function of perpendicular magnetic field at zero asymmetry \cite{01Bfuhrer} is depicted in Fig.~\ref{fig1}(d). At B=77~mT and $\alpha=0$ (black horizontal lines) the two spectra are identical except near the points where a charge rearrangement (black dashed lines) adds additional offsets to some peak positions. The amplitudes of the peaks (not shown) for the two sweeps show the same magnitudes. We conclude that the ring is in the same state for both measurements taken in the same cooldown. 

In a constant interaction picture spin pairs are expected to occur with a parallel evolution as a function of any parameter tuning the orbital part of the wave function. In contrast, Fig.~\ref{fig1}(c) shows peaks changing their spacing as a function of $\alpha$.  For each peak linear sections are observed belonging to one of two classes of slopes $a$ and $b$ (shown in the inset). 
At the transition between two linear parts a cusp occurs [black arrow in Fig.~\ref{fig1}(c)] which usually has a counterpart in one of the neighboring peak positions. 
We argue in the following that these features are due to level crossings between states $b$ with a probability density extended around the ring and states $a$ that are rather localized in one arm of the ring.

We denote states with a probability density extending around the ring as `extended'. In the same spirit a `localized' state connects to both, source and drain, but has appreciable probability density predominantly in one arm of the ring.
Extended states are strongly influenced by AB-flux through the central annulus. They exhibit a strong zig-zag motion with a period $\Delta B =75$~mT corresponding to the addition of one flux quantum. This magnetic finger print is due to alternating crossings of well defined angular momentum states~\cite{01Bfuhrer}. These levels are experimentally found to occur in spin pairs [peaks~3 and 4 as well as 9 and 10 in Fig.~\ref{fig1}(d)], i.e. the orbital levels are successively occupied with  spin up and spin down. In a sequence of 16 peaks we find more than 12 pairwise correlated suggesting a comparatively small influence of interactions beyond the constant interaction model.

However, some peaks, e.g. peak 2 in Fig.~\ref{fig1}(d), do not exhibit such pairing. We attribute this behavior to the more localized character of the wave function. In a realistic ring structure perfect rotational symmetry may always be slightly perturbed by source and drain junctions, a small geometric asymmetry or background potential fluctuations. These perturbations lift degeneracies in the ring spectrum~\cite{84buttiker} and mix angular momentum states at small energies in each radial subband. A sufficiently strong asymmetry may lead to localization of some states in one arm of the ring and a flat magnetic field dispersion is expected.
The interaction energy necessary to charge a second electron into one of these states is larger than that for extended states and spin pairing is less likely to be observed.

Measurements as a function of asymmetry [Fig.~\ref{fig1}(c)] help to identify localized states. States which are more localized in the upper arm of the ring will tend to lower their energy with increasing asymmetry, whereas those in the lower arm will be raised. For a rotationally symmetric state extended around the ring these two effects cancel and in first order the level is independent of asymmetry in analogy with the Stark effect.

We identify the levels with a steep positive slope as a function of $\alpha$, i.e. peak 2 (purple) and 5 (red) at $\alpha=0$ in Fig.~\ref{fig1}(c), with states that are more localized in the lower arm of the ring. The evolution of these levels is indicated by the dashed sections of the peak positions. Accordingly, they show a rather flat $B$-dispersion in Fig.~\ref{fig1}(d). While this is obvious for peak 2 (purple) it is not immediately clear for peak 5 (red) which shows a zig-zag behavior as a function of magnetic field. However, looking carefully at the evolution of this level at $B=77$~mT we find a short flat section. Together with the absence of spin pairing, we conclude that there is a localized state close to an extended state in agreement with the asymmetry data,
which frequently cross as a function of $B$.

The clearest case of an extended state is observed for peaks 3 and 4 (cyan and green). Here, we observe correlated zig-zag behavior over a large $B$-range as well as a small slope of both peaks in a large interval of $\alpha$. They run in parallel until a localized state crosses indicating identical orbital wavefunctions and spin pairing. 

In the following we discuss such a crossing
between a localized and an extended spin pair in a Hartree-Fock picture, which allows us to extract Hartree and exchange contributions to the interaction from our data.

\begin{figure}[tbp]
\centering
\includegraphics[width=3in]{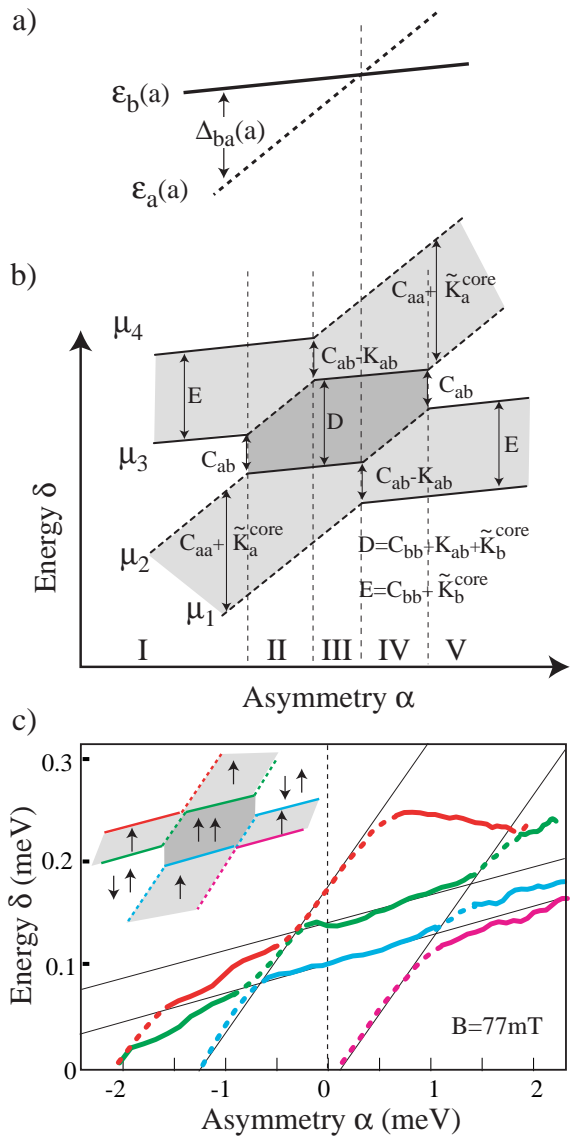}
  \caption{\small (a) Schematic representation of two orbital levels $a$ and $b$ that cross as a function of asymmetry $\alpha$. (b) Expected behavior for the electrochemical potentials of the ring with N+1 to N+4 electrons. The arrows indicate conductance peak spacings due to interaction contributions. The experimental constant $E_{\small C}$ can be identified with the energy $C_{\mathrm{ab}}-K_{\mathrm{ab}}$, i.e. the energy needed to charge orbital $a$ in the presence of an electron with parallel spin in orbital $b$. The roman numbers indicate characteristic ranges of $\alpha$. (c) Inset: spin occupation of the two uppermost levels close to an orbital crossing as a function of asymmetry and energy. The dark rhomboid region in the center indicates the spin-triplet region. Main figure: experimental realization of such a situation.}\label{fig2}
\end{figure}

To this end we follow the notation of Ref.~\onlinecite{00tarucha} where the uppermost levels of a few electron dot were studied. In contrast to their model we do not neglect the interaction with lower lying (core) levels since we do not know the spin configuration of the dot.
We neglect self-consistent changes of the single-particle orbital wavefunctions~\cite{34koopmans} with changing number of electrons from N+1 to N+4.
 This is generally taken to be a good approximation for regular many-electron dots with $N \gg 1$ and weak interactions~\cite{99walker}. 
The electrochemical potential of the ring is then given by~\cite{93schwabel}
\begin{equation}
\mu_{N}\left(\alpha\right)=\epsilon_{N}(\alpha)+
\sum_{j=1}^{N-1}\left[C_{Nj}-\delta_{s_N s_j}K_{Nj}\right]
\label{themodel}
\end{equation}
where the overlap integrals for the Hartree and the exchange terms are 
\begin{eqnarray}
C_{ij}&=&e^2\int d\vec{r}d\vec{r}'|\phi_i(\vec{r})|^2G(\vec{r},\vec{r}')|\phi_j(\vec{r}')|^2\nonumber\\
K_{ij}&=&e^2\int d\vec{r}d\vec{r}'\phi_i^*(\vec{r})\phi_j^*(\vec{r}')G(\vec{r},\vec{r}')\phi_i(\vec{r}')\phi_j(\vec{r})\nonumber
\end{eqnarray}
$s_i$ is the spin associated with the single-particle wavefunction $\phi_i$ and the index $i$ runs over particles on the ring counting each filled orbital twice. The function $G(\vec{r},\vec{r}')$ is the electrostatic Green's function of the system.

Fig.~\ref{fig2}(a) shows a crossing of the two uppermost orbital single-particle energy levels as a function of $\alpha$. Level $\epsilon_b(\alpha)$ (bold line) is an extended state,  $\epsilon_a(\alpha)$ (dashed line) a more localized state and the single-particle level spacing $\Delta_{ba}$ depends on $\alpha$. Below these single particle levels we consider $N$ electrons occupying core states of the ring. In Eq.~(\ref{themodel}) we split the interaction term into a part due to core levels and a part due to levels $a$ and $b$. Assuming that the core electron configuration does not change when levels $a$ and $b$ are filled, the Hartree contribution of the core levels is irrelevant for the peak spacings considered. The exchange contribution of the core levels, however, depends on the spin which is being filled into the ring and is given by
$\tilde{K}_{i}^\mathrm{core}=\sum_{j<a,b} \left(\delta_{\uparrow s_{j}} -\delta_{\downarrow s_{j}} \right) K_{ij}$.
The interactions of levels $a$ and $b$ with the lower levels can only be neglected if each core orbital is doubly occupied resulting in total core spin 0.

Fig.~\ref{fig2}(b) shows the expected peak positions and spacings schematically.
The first additional electron will occupy the lower of the two orbitals $a$ and $b$ ($\mu_1$). 
Far from the crossing [regions I and V in Fig.~\ref{fig2}(b)] the second electron ($\mu_2$) will fill the same orbital and form a singlet with the electron which is present. The level spacing $\mu_2-\mu_1$ is given by the Hartree interaction $C_{ii}$ between the two electrons on the same orbital plus an exchange contribution $\tilde{K}_{i}^\mathrm{core}$.  The Hartree contribution is larger for a localized state ($C_{aa}>C_{bb}$) and the separation between the peak spacings in region I is therefore larger than in region V.

Closer to the crossing we expect a singlet-triplet transition regarding  the total spin of the two uppermost levels as a function of $\alpha$ when the gain in exchange energy for parallel spins becomes larger than $\Delta_{ba}$. This leads to kinks in the peak position $\mu_2$ in Fig.~\ref{fig2}(b)~\cite{00baranger}.
However, in the case of two levels with different probability densities fluctuations in the Hartree term contribute as well. In our case the fact that $C_{ab}<C_{aa}$ supports the exchange interaction in shifting the transition $I\rightarrow II$ to the triplet state to smaller $\alpha$. This leads to kinks in $\mu_2$ even if the exchange contribution is negligible (e.g. because it is smaller than the thermal energy). The pairs of kinks in $\mu_2$ and $\mu_3$ at the transitions from region $I\rightarrow II$ and $IV \rightarrow V$ always occur at the same asymmetry value. The extent of region III depends on the difference $C_{bb}-C_{aa}$ and vanishes only for levels with the same probability densities.

Exchange and Hartree contributions can be discerned by following the evolution of a spin pair as a function of $\alpha$. For small $\alpha$ (region I) the separation between $\mu_3$ and $\mu_4$ is given 
by $E=C_{bb}+\tilde{K}^\mathrm{core}_{b}$. In the triplet region III the separation $D=\mu_3-\mu_2$ is larger by the exchange energy $K_{\mathrm{ab}}$ between the two orbitals. For large $\alpha$ (region V), $E$ is recovered in $\mu_2-\mu_1$. We show in the following that this pattern is observed in the experiment.

In Fig.~\ref{fig2}(c) we focus on peaks~3 (cyan) and 4 (green) which we have identified as a spin pair of an extended state for small $\alpha$ and two more localized levels (magenta and red). The separation between the two peak positions changes whenever a localized level crosses, i.e. peaks~3 and 4 at $\alpha=0$~meV (separation $D$) are further apart than peaks~4 and 5 at $\alpha=-1.5$~meV (separation $E$) and peaks~2 and 3 at $\alpha=+1.5$~meV (separation $E$).
Comparing these variations with the model (Fig.~\ref{fig2}(b) and (c) inset) we identify the central region with the triplet state. Increasing $\alpha$ while staying between peak~2 and 3 tunes the ring from a singlet to a triplet back to a singlet state. 

Analyzing the peak separations with the model we determine the interaction contributions given in table~\ref{tab1}.
\begin{table}[b]
\begin{center}
\begin{tabular}{|c|c|c|c|}
\hline
  $C_{aa}$ & $C_{bb}$ & $C_{ab}$ & $K_{ab}$\\ \hline
  $530\mu$eV &  $360\mu$eV  &  $335\mu$eV &  $25\mu$eV \\
\hline
\end{tabular}
\end{center}
  \caption{\small Interaction contributions extracted from the data.}\label{tab1}
\end{table}
The uncertainty of these values is of the order of $10\mu$eV comparable to the thermal energy. The value of the exchange contribution is small but larger than the thermal broadening of the peaks. The value for $C_{aa}$ was estimated by assuming that the two localized states indicated by the steep thin black lines in Fig.~\ref{fig2}(c) belong to the same orbital wave function.
We expect that any localized orbital wave function has to fill at least one arm of the ring, given that a conductance peak is observable. This gives an upper bound of two for the ratio $C_{aa}/C_{bb}$, in agreement with the analysis. Additional measurements taken during a second cooldown in a regime where the ring was more strongly coupled to the contacts confirm the general findings for the crossings of localized and extended states and the corresponding kinks in the peak positions. However, in this case we were not able to resolve exchange effects due to a larger peak width.

Our results show that in a Coulomb-blockaded many-electron ring (dot) the gate voltage controlled preparation of a specific spin state is possible at low temperatures. The special arrangement with a separate gate for each arm of the ring allowed to tune the ring's symmetry. We have demonstrated an asymmetry induced singlet-triplet transition. Energy level shifts as a function of asymmetric gate voltages and magnetic field were related to properties of the corresponding wavefunctions. Hartree and exchange interaction contributions were quantitatively determined.

\end{document}